\documentclass[useAMS,amssymb,epsfig,usegraphicx,usenatbib,twocolumn,revtex4-1]{emulateapj}

\def\aj{AJ}

\def\apj{ApJ}
\def\apjl{ApJ}
\def\apjs{ApJS}
\def\aap{A\&A}
\def\mnras{MNRAS}
\def\na{NewA}
\def\pasp{PASP} 

\def\physrep{Physics Reports}

\usepackage[usenames,dvips]{color}

\newif\ifAMStwofonts

\makeatother

\shorttitle{Long-Lived Spiral Structures}
\shortauthors{Saha and Elmegreen}
\begin{document}
\title{Long-Lived Spiral Structure for Galaxies with Intermediate Size Bulges}

\author {Kanak Saha$^{1}$, \& Bruce Elmegreen$^{2}$}
\affil{$^{1}$ Inter-University Centre for Astronomy and Astrophysics, Pune 411007, India, \\
$^{2}$ IBM Research Division, T. J. Watson Research Center, 1101 Kitchawan Road, Yorktown
Heights, NY 10598, USA
\\e-mail: kanak@iucaa.in, bge@us.ibm.com}

\label{firstpage}

\begin{abstract}
Spiral structure in disk galaxies is modeled with nine collisionless N-body simulations
including live disks, halos, and bulges with a range of masses.  Two of these simulations
make long-lasting and strong two-arm spiral wave modes that last for $\sim5$ Gyr
with constant pattern speed. These two had a light stellar disk and the
largest values of the Toomre $Q$ parameter in the inner region at the time the spirals
formed, suggesting the presence of a Q-barrier to wave propagation resulting from the
bulge. The relative bulge mass in these cases is about 10\%. Models with 
weak two-arm spirals had pattern speeds that followed the radial dependence of the Inner
Lindblad Resonance.
\end{abstract}

\keywords{galaxies: bulges --- galaxies: evolution --- galaxies: kinematics and dynamics
--- galaxies: spiral --- galaxies: structure}

\section{Introduction}
\label{sec:intro}
Spiral structure in disk galaxies results from gravitationally amplified growth of
density perturbations (\citealt{GoldreichLyndenBell65,Carlberg85,ToomreKalnajs91, Huber02}, see
reviews in \citealt{Athanassoula84,Bertin96}) that range in scale from
interstellar clouds \citep{DOnghiaetal2013}, to other spirals \citep{Masset1997}, to bars
\citep{Salo10} and passing galaxies \citep{ToomreToomre72,Salo00}. Simulated spiral arms
are usually individually short-lived (\citealt{Sellwood2011}; however, see
\citealt{Elmegreen1993}), although the presence of spiral structure in one form or
another may be long-lived \citep{Sellwood2014}.

\cite{Lindblad1962,Lindblad1963} considered spirals ``quasi-stationary'' when he proposed
they might result from synchronized epicyclic motions over a wide range of radii as
viewed in a rotating coordinate system. This was a better model than material arms which
would wrap up too quickly to explain the high fraction of disk galaxies with spirals.
However, the correspondence noted by Lindblad was not perfect and the proposed arms would
still deform over time. \cite{LinShu1964} offered a solution to this problem by noting
that arm self-gravity would perturb the epicycles in the right sense and cause them to
lock into phase. Their solution was only a start, though, as \cite{Toomre1969}
pointed out that Lin-Shu waves also wrap up because their group velocity is inward.

The second part of this solution proposed that incoming waves refract or reflect in the
central regions and move back out \citep{Lin70,Mark1976a}. Then they can amplify by disk
gravity at corotation (CR) and send another trailing wave back in as well as a trailing
wave out \citep{Mark1976b,Mark1976c}.  The result is a growing spiral wave mode, which is
a standing wave with components moving in both directions
\citep{Mark1977,Bertin1989a,Bertin1989b}. In the WASER type II mode discussed by
\cite{Bertin1983} and \cite{LinBertin1985} an inward-moving trailing spiral wave
reflects off a high-Q barrier in the inner region, such as a bulge, and returns to CR as
a weak leading wave. The leading wave then swings around into a stronger trailing
wave in analogy to the swing amplifier proposed by \cite{Toomre1981}, and the trailing
wave moves inward again.  The outward moving trailing wave beyond CR extends to the outer
Lindblad resonance where it resonates with the thermal motions of stars.  For such a wave
mode, the pattern speed is determined by the propagation condition for a wave to move in
to the reflection radius and back out to CR in the time it takes the pattern to rotate to
another arm \citep{Bertin1989b}.

Astronomical evidence for long-lived global modes is weak. \cite{EE1983} suggested that
two-arm spirals live for at least $\sim2$ Gyr considering an observed increase in the
fraction of such ``grand-designs'' with galaxy group crossing rate and galaxy-galaxy
collision rate. This duration corresponds to $\sim5$ rotations in the outer parts of a
typical galaxy and is reasonably consistent with a spiral mode. Morphological evidence
for spiral modes was suggested in \cite{Elmegreen1989} and \cite{Puerari2000} by the
presence of symmetric spiral arm amplitude variations from interfering inward and outward
moving waves \citep[see also][]{Bertin1993}. The arm variations observed for M81 were fit
to the modal theory by \cite{Lowe1994}. Further evidence for modes was shown in
\cite{Elmegreen1992} by the presence of 2, 3, and 4-arm symmetric spirals in 18 galaxies,
with regular amplitude variations along the arms and arm endpoints at the appropriate
Lindblad resonances. These latter two studies used computer-enhanced and symmetrized
images to show features that are not obvious to the eye.

Modern simulations always have transient spiral features, even when they are wave modes
\citep{Sellwood2011}. This transience is partly because the modes adjust the radial
distribution of stars and their velocity dispersions to change the basic state
\citep{Sellwood2012}, and they also trigger new modes which spring up at resonances of
the old ones \citep{Sygnet1988}. Still, one wonders if a simulation with the right
initial conditions can make a single, long-lasting wave mode in the sense discussed by
\cite{Bertin1989b} and others. One essential component is a bulge that shields the inner
Lindblad resonance (ILR) from incoming waves so they can reflect back to the
amplification zone at CR. Otherwise the ILR will absorb the wave
\citep{Lynden-BellKalnajs1972}.

The present paper shows several examples of particle simulations that have a live disk,
bulge, and halo and that appear to make single, long-lasting wave modes with constant
pattern speeds. They have the strongest amplitude when the bulge is optimal for shielding
the ILR.  The models are in a sequence of increasing bulge mass, which also increases the
strength of the ILR. For low-mass bulges, the ILR is weak but the bulges are also weak,
and the spirals end up weak themselves. For high-mass bulges, the ILR absorption and
bulge shielding are both strong but the ILR apparently wins again, making the waves
weak. But for intermediate mass bulges, the conditions are optimal to make a strong and
persistent two-arm wavemode, which circulates at constant pattern speed from $\sim 4$ Gyr
to $\sim8$ Gyr in the simulation. Eventually the large arm amplitudes that result from
this continuous growth can distort and destroy the spiral mode.

\begin{figure}
\begin{flushleft}
\rotatebox{0}{\includegraphics[height=6.3 cm]{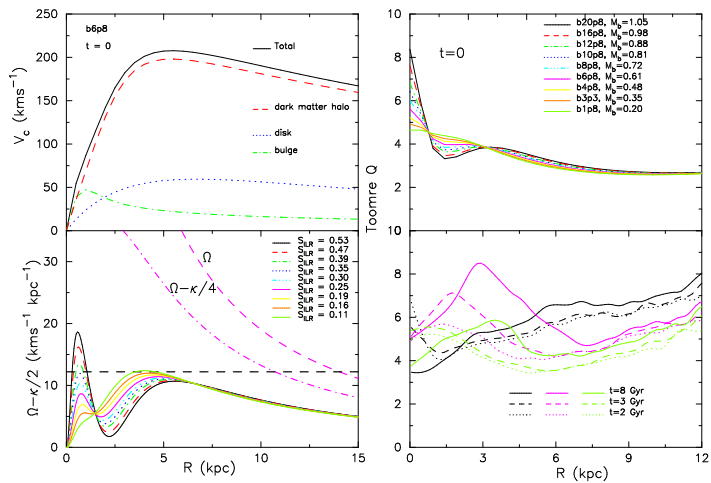}}
\caption{Initial set up and Q-bump during the growth of spiral arms.
Upper panel(left): circular velocity curves. Lower panel(left):
radial variation of $\Omega, \Omega - \kappa/2,\Omega - \kappa/4 $ for
different model galaxies. $S_{\rm ILR}$ denotes the ILR strength defined in
section~\ref{sec:simulation}.
Upper panel (right): Initial Toomre Q profiles for all the model galaxies.
The bulge mass ($M_b$) is in units of $10^{9}M_{\odot}$. Lower panel(right):
The Toomre Q parameter versus radius for models b20p8(black), b6p8 (magenta) and b1p8 (green).
 Q for model b6p8 develops a peak in the inner region over time, and grows strong spiral arms
when the peak is largest (Fig.~\ref{fig:A2rt}). Model b1p8 has a smaller Q peak later
and grows a weak spiral later.
Model b20p8 has no significant Q peak and very weak arms throughout the simulation.
Only the models with a Q peak
in the inner parts developed strong two-arm spirals, and they did this after the
Q peak appeared.}
\label{fig:omgkapa}
\end{flushleft}
\end{figure}

\section{Model set up and simulation}
\label{sec:simulation} We have constructed a set of nine initially equilibrium models of
disk galaxies using the self-consistent method of \cite{KD1995}. Each galaxy model
consists of a stellar disk with an exponentially falling surface density, a flattened,
cored dark-matter halo, and a classical bulge modeled with a King DF \citep[for
details see][]{sahaetal2012}. All three components are live. In order to understand the
role of the ILR, the disk and dark matter halo parameters are kept same in all
models, and only the total mass of the bulge ($M_b$) varies. This allows us to
investigate the growth of spiral structure as a function of the bulge-to-total mass ratio
($B/T$). The initial variation of $\Omega - \kappa/2$ with radius is shown in the bottom
left of Figure~\ref{fig:omgkapa} for all models. The models are identified by an
ILR strength, $S_{\rm ILR}$, which is defined to be the ratio of the peak value of
$\Omega-\kappa/2$ in the inner parts to the rotation rate at $2$ scale lengths,
$\Omega(2R_{\rm d})$. The ILR for the $m=2$ pattern occurs where the pattern speed equals
$\Omega -\kappa/2$; the higher the bump at small $\Omega - \kappa/2$, the more likely
there will be an ILR for the pattern speed chosen by the mass distribution.

\begin{figure}
\rotatebox{0}{\includegraphics[height=11.0 cm]{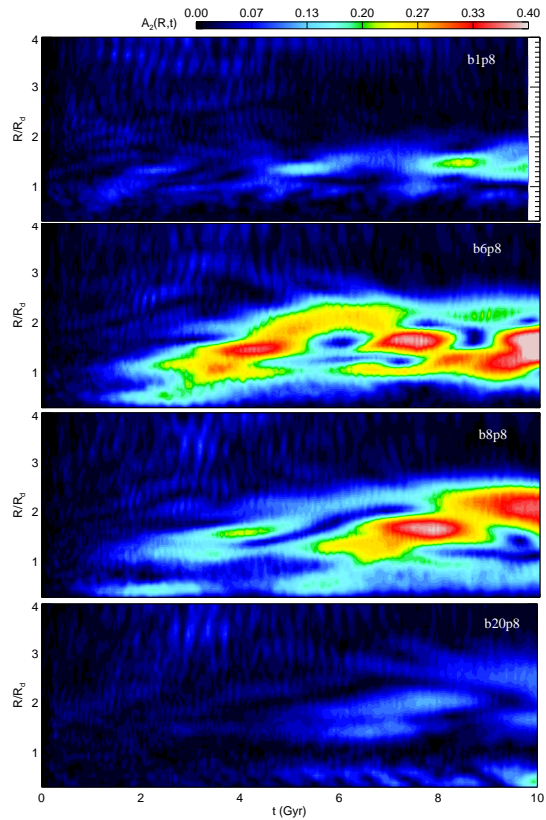}}
\caption{Spatio-temporal variation of the $m=2$ Fourier amplitude in four models; 
$R_d=3 kpc$. Model b20p8 has essentially no spiral structure.}
\label{fig:A2rt}
\end{figure}

\begin{figure*}
\rotatebox{0}{\includegraphics[height=8.0 cm]{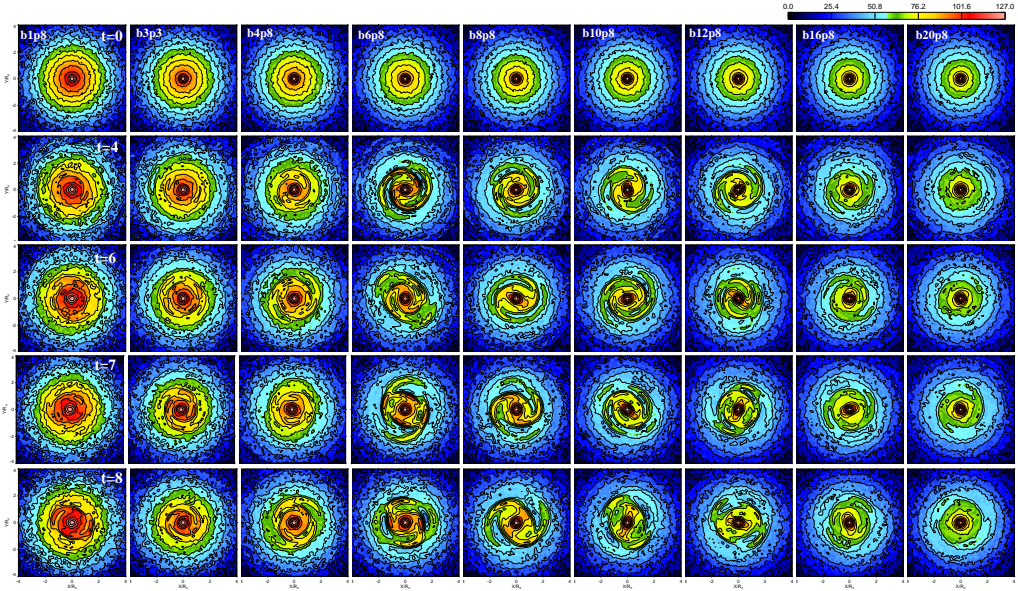}}
\caption{Face-on surface density maps for each model and their evolution. Time increases
downward (in units of Gyr) and bulge mass increases from left to right. The labels for
the models indicated at the top of each column correspond to bulge densities, as
discussed in the text. The color scale at the top right corresponds to relative surface
density. The leading spiral at $t=7$ for b6p8 is the beginning of the outer ring visible
at $t=8$.} \label{fig:surfden}
\end{figure*}

The model number indicates the bulge central density ($\rho_b$) which is one of the parameters
of a truncated King model, e.g., b10p8 corresponds to a $\rho_b = 10.8$ (internal units). 
Higher values of $\rho_b$ correspond to higher bulge masses. Each model is scaled such that 
the scale length is $R_{\rm d} = 3$~kpc and the circular speed $V_c = 200$ km s$^{-1}$ at 
$2 R_{\rm d}$ (see Fig.~\ref{fig:omgkapa}). The ratio of disk-to-halo mass 
$M_d/M_h = 0.07$ for all models - indicating a light stellar disk.
For model b6p8, $M_d = 6.6 \times 10^{9} M_{\odot}$ and $M_h = 9.3 \times 10^{10}
M_{\odot}$. The bulge mass varies from $0.2 \times10^{9} M_{\odot}$ for b1p8 to $1.0
\times 10^{9} M_{\odot}$ in b20p8; see Figure~\ref{fig:omgkapa}.

The initial stellar disk is relatively hot with Toomre $Q(R) \equiv \frac{\kappa \sigma_{\rm R}}{3.36 G \Sigma}$ varying across the disk (top right panel of 
Fig.~\ref{fig:omgkapa}). Q is relatively high because the stellar disk is comparatively 
low-density. For example, model b6p8 has a central surface density of 
$\Sigma_{0} = 117.4\;M_{\odot}$ pc$^{-2}$, which can easily be obtained from the
total mass and length scale given above. This is about 5 times lower than that of Milky
Way's stellar disk \citep{KuijkenGilmore1991}. The radial velocity dispersion
($\sigma_{\rm R}$) falls off also exponentially with twice the scale length of the disk
to keep the initial scale height constant at $300$~pc for our stellar disk
\citep{LewisFreeman1989,KD1995}. The stars in the disk move under the 
epicyclic approximation (as par model construction), according to which the 
ratio of azimuthal to radial velocity dispersion follows the relation \citep{BT1987}:

\begin{equation}
\frac{\sigma_{\rm \varphi}}{\sigma_{\rm R}} = \frac{\kappa}{2 \Omega}.
\end{equation}

\noindent We have checked this for all the models at different times and found that the
above relation holds pretty well. For example, for model b6p8, at $t=2$ Gyr, at $2R_{\rm
d}$, $\kappa=45.53$ km s$^{-1}$ kpc$^{-1}$ and $\Omega=32.65$ km s$^{-1}$ kpc$^{-1}$
giving $\kappa/{2\Omega}=0.697$. While $\sigma_{\rm R}=19.95$ km s$^{-1}$ and
$\sigma_{\rm \varphi}=14.02$ km s$^{-1}$, giving $\sigma_{\rm \varphi}/\sigma_{\rm
R}=0.7$ in good agreement. Note that $Q(R)$ varies with time as well, as the disk mass
and stellar velocity dispersion change due to spiral forcing and heating
\citep{JenkinsBinney1990,SellwoodBinney2002, Sahaetal2010,Roskeretal2012}. $Q(R)$ is
shown at $3$ times during the simulation in the bottom right of
Figure~\ref{fig:omgkapa}. The model with the highest peak in $Q(R)$, model b6p8 (magenta
curve, $M_b=0.61$), also has the strongest spiral structure at that time (see below).

\begin{figure}
\rotatebox{0}{\includegraphics[height=7.0 cm]{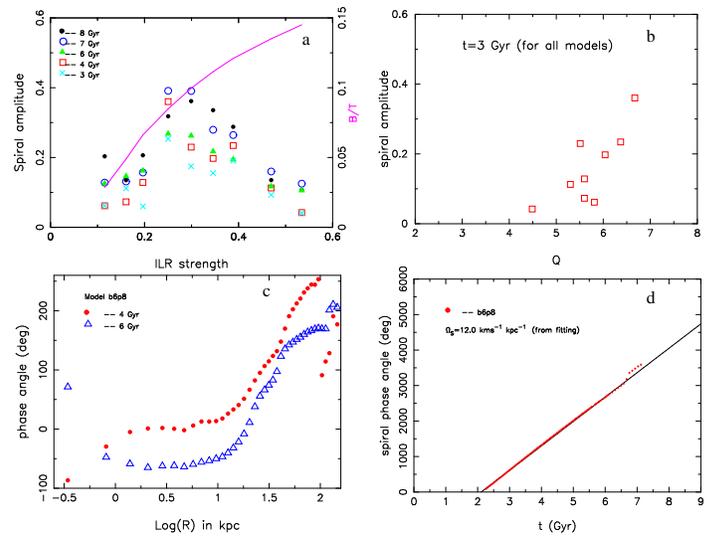}}
\caption{(a) Dependence of the Fourier $m=2$ amplitude of the spiral 
pattern at the indicated times on ILR strength (defined in sec.~\ref{sec:simulation}); 
pink solid line shows the dependence of $B/T$ on the ILR strength. (b) The same on 
the peak value of Q computed at $3$~Gyr. (c) Phase angle ($m=2$) versus
natural log of radius in kpc. (d) The temporal variation of the phase angle at a given 
radius for the spiral mode run b6p8. The solid line is the linear fit to the phase angle
versus time; the slope is the pattern speed of $12.0$~kms$^{-1}kpc^{-1}$ }
\label{fig:A2BT}
\end{figure}

\begin{figure*}
\rotatebox{0}{\includegraphics[height=7.9 cm]{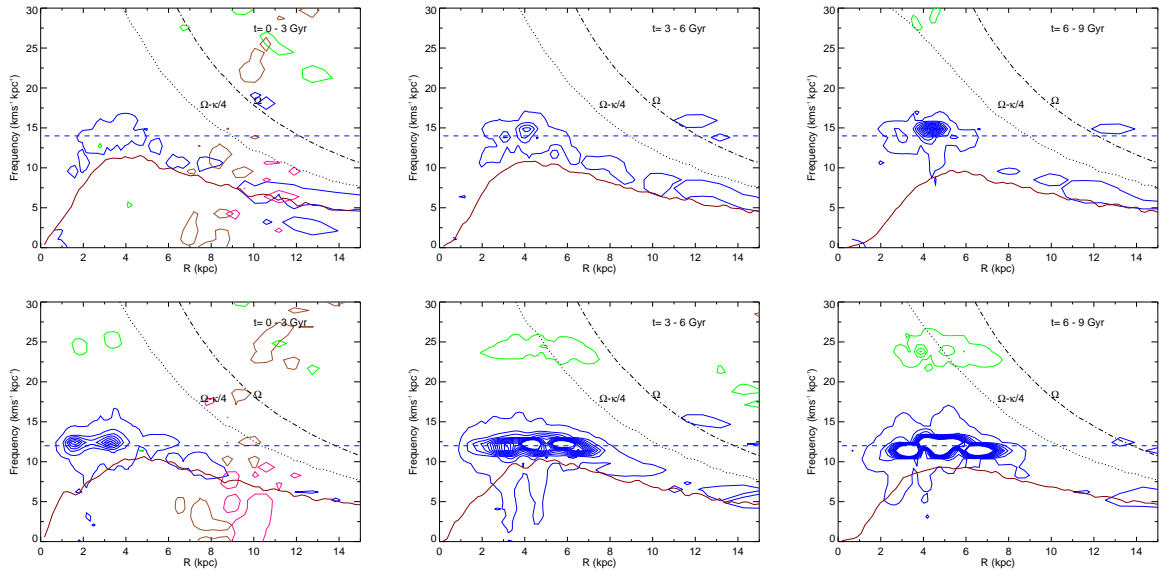}}
\caption{Basic dynamical frequencies and associated resonance locations at
three different time intervals for models b1p8 (upper panels) and b6p8 (lower panels).
Overplotted on this is the power spectrum for the $m=2$ (in blue), $m=4$ (green),$m=6$
(brown), and $m=8$ (pink) Fourier mode, computed at the indicated the time interval. The
peak value of the power spectrum is $4\times 10^{-3}$, while the outermost is at $2\times
10^{-5}$. The strongest part of the spiral arms as shown by the contours ends in the
inner region at the position of the Q-barrier, shown in the lower right panel of
Fig.~\ref{fig:omgkapa}.} \label{fig:phase}
\end{figure*}

We used a total of $2.2 \times 10^6$ particles to simulate each model galaxy of which
$1.0 \times 10^5$ were in the classical bulge, $1.05 \times 10^6$ in the disk and $1.05
\times 10^6$ in the dark matter halo. The time unit for model b6p8 is $45.3$~Myr.
The softening lengths for the disk, bulge and halo
were unequal and chosen so that the maximum force from particles of all species (bulge,
disk, halo) is nearly the same \citep{McMillan2007}. The simulations were performed with
the Gadget code where forces between the particles were computed using a modified 
Barnes \& Hut tree with a tolerance parameter $\theta_{tol} = 0.7$ \citep{Springeletal2001}. 

\section{Growth of spiral structures}
\label{sec:spirals} In order to characterize the growth of spiral structure in the
simulated stellar disks, we compute the $m=2$ Fourier component of the disk
particle density and monitor its amplitude and phase variation across the disk as well as
a function of time elapsed in the simulation. The development of spiral arms as a
function of radius and time is shown in Figure~\ref{fig:A2rt}, which plots the two-arm
Fourier amplitude, $A_2(R,t)$, as a color scale for two strong (b6p8, b8p8) and
two weak (b1p8, b20p8) spiral models. When there are $m=2$ spirals, the amplitudes vary
with time in pulses roughly separated by $\sim3$ Gyr at $R/R_{\rm d}\sim 1.5 - 2$, where
the rotation time is $\sim 0.16$ Gyr. The strongest two-arm spiral structure
appears in model b6p8 hosting an intermediate bulge with $B/T=0.085$; initially 
(upto about $4$~Gyr) the spiral grows exponentially with a growth rate 
$\omega_I= 0.63$~km s$^{-1}$kpc$^{-1}$; after which the amplitude growth slows down 
with $\omega_I= 0.082$~km s$^{-1}$kpc$^{-1}$ - showing signs of saturation. 
We also re-ran the model b6p8 with $6.15 \times 10^{6}$ particles 
(with $5$ million in the halo) to check 
the effect of halo shot noise and found the primary results e.g., pattern speed, 
spiral amplitude unchanged; but the growth of the spiral is delayed by about $1.4$~Gyr
with nearly the same growth rate $\omega_I = 0.66$~km s$^{-1}$kpc$^{-1}$.

Figure~\ref{fig:A2rt} also shows that at $\sim 3$~Gyr, the spiral amplitude in model b6p8
faces a sudden growth. This growth is associated with a peak in the Toomre Q profile (see
Fig.~\ref{fig:omgkapa}), which we interpret as a Q-barrier to the inward propagation of
spiral arms. No other models had such a large Q-peak at that early time, but all of the
models with large spiral amplitudes had a prominent Q-peak just before the spirals
developed. Model b8p8 (with $B/T=0.1$, Fig.~\ref{fig:A2rt}) grows strong two-arm
spiral structure between 6 and 7 Gyr, and has a strong Q-peak at that time. The same is
true for model b1p8: although it formed only a weak spiral with no bar
(Fig.~\ref{fig:A2rt}), after a weak inner Q-peak appeared (Fig.~\ref{fig:omgkapa}) 
at $\sim 8$~Gyr. The simultaneous occurrence of spiral structure and
Q-peaks suggests some feedback effect in which an initially weak spiral changes the disk
to produce a Q-peak, which, in turn, makes the spiral stronger. The model with a very
high mass bulge, b20p8, evolved without growing either a Q-peak or a significant spiral.

Figure~\ref{fig:surfden} shows the face-on surface density maps for all of the 
models at times of $0, 4, 6, 7, 8$~Gyr. Model b1p8 on the left has the lowest mass 
bulge and evolves rather passively without making strong spirals until about 
$8$~Gyr. As the bulge mass increases (moving to the right in the figure), the spiral 
structure becomes more prominent. As mentioned above, model b6p8 forms a strong, 
two-arm spiral at $\sim 3$~Gyr and maintains that for the next $\sim 4$~Gyrs. 
Interestingly, just before $2$~Gyr in this model, a weak and short ($<1.5$~kpc) 
bar-like structure that rotates with a different pattern speed, forms within the 
ILR. Images made from the disk particles show that this $m=2$ feature resembles 
a nuclear bar; it disappears after $\sim 4$ Gyr. Model b8p8 has an ILR strength 
slightly higher than that of b6p8 and the spiral structure in
the initial phase is weaker, but it progressively becomes stronger and develops a
bar-like feature eventually. b10p8 also develops spiral structure but now there is a 
clear bar; the spiral dissolves gradually by $8$~Gyr and the bar grows stronger. 
Model b12p8 has an even stronger ILR and forms a comparatively weaker spiral structure 
associated with a bar. The other two models, b16p8 and b20p8, have massive bulges and 
higher ILR strengths but do not form any conspicuous spiral structure.
It is interesting to note that both b1p8 and b20p8 did not grow a bar even after many
tens of rotation times. Although b1p8 had no ILR, the disk was quite hot ($Q
> 4$) and associated with a high value of the swing-amplification parameter
\citep{BT1987}, $X = 14$ at $R = R_{\rm d}$, making the
swing-amplification process ineffective.

Figure~\ref{fig:A2BT} (a,b) together suggest that the strongest spiral 
structure forms at intermediate ILR strengths and relative bulge masses, when the 
peak in $Q(R)$ is large. We interpret this Q-peak as a measure of the $Q$-barrier 
to the ILR. The radial variation of the phase angle at $4$ and $6$ Gyr suggest the 
presence of a bar-like structure which roates with a pattern speed same as the 
spiral in b6p8. This bar is also seen Fig.~\ref{fig:A2rt}
as a non-zero Fourier amplitude within $1 R_d$ at early times, but it weakens at 
later times (Fig.~\ref{fig:A2rt}) as may also be seen from the inner nearly circular 
density contours at late times in Fig.~\ref{fig:surfden}.

\subsection{Wavy nature}
It remains to verify whether these spiral structures are density wave modes as proposed
by Lin, Mark, Bertin and others. Fig.~\ref{fig:A2BT}(d) shows the phase
angle versus time for the $m=2$ Fourier component of the spiral in run b6p8, which has
the strongest arms at 4 Gyr. The phase angle was calculated in a radial range around the
peak of the spiral amplitude, $R/R_{\rm d}=1.2$. The phase increases regularly with time,
indicating a constant pattern speed equal to the slope, $\Omega_{\rm p}=12.0$ km
s$^{-1}$~kpc$^{-1}$ from 2 Gyr to 7 Gyr. This pattern speed is shown by a dashed line in
the lower panels of Figure~\ref{fig:phase}, along with the other angular 
rates indicated. For each time interval, the $m=2$ (blue), $m=4$ (green) and 
$m=6$ (brown),$m=8$ (pink) Fourier power spectra for b6p8 were computed directly from the particle 
distributions \citep[following][]{Sellwood1985,Masset1997} and overplotted as 
contours. The peak of the blue contours indicates a coherent spiral 
with a long-lasting pattern speed of $\sim12.0$~km s$^{-1}$kpc$^{-1}$, 
close to the peak of the ILR curve. The corotation resonance (CR) is where 
$\Omega_{\rm p}=\Omega(R)$ and lies at $R_{\rm cr} = 13.5$~kpc from the center. 
The strong spiral in model b8p8 also has a constant pattern speed, 
$\Omega_{\rm p}=11.9$ km s$^{-1}$ kpc$^{-1}$, which lasts for about $4.5$~Gyr.

Higher order waves have higher pattern speeds. The $4:1$ resonance for 
the $m=2$ pattern speed of model b6p8 is at $10.5$~kpc ($3.5 R_{\rm d}$) in Figure
~\ref{fig:phase}. A faint $m=4$ spiral lies on that resonance at high frequency. There
are more high-order waves at earlier times when the spirals are weak. The weak-spiral
model b1p8 (Fig.~\ref{fig:phase}, top) also has high-order waves at early times, along
with weak $m=2$ waves. These weak two-arm spirals follow the ILR with a pattern speed
that varies with radius, reminiscent of Lindblad spirals caused by in-phase epicycles
with weak gravity till about 8 Gyr, after which it rotates with a pattern speed 
$\Omega_p \simeq 14$km s$^{-1}$ kpc$^{-1}$. 
Generally, the strong spirals in our models stay within the $4:1$ resonance, in 
compliance with K-band observations by \cite{GrosbolPatsis1998}.

\section{Conclusions}
\label{sec:discuss} Two of our particle simulations with a live halo, bulge and 
a light disk were found to generate spiral wave modes that grew slowly 
at first over a period of 1 to 2 Gyr
and then quickly to a relatively large amplitude over the next 1 Gyr, after which they
maintained a constant pattern speed for another 5 Gyr. These two simulations differed
from the others, which did not make strong spiral arms, in the relative height of the
maximum value of the Toomre Q parameter in the inner region. This inner Q peak results
from the bulge. We interpret this result as evidence that wave reflection off a classical
bulge can lead to the formation of a long-lasting spiral wave mode, as proposed by 
\cite{Bertin1989a}.

\medskip
\noindent{\bf Acknowledgement:} The authors thanks the anonymous referee for several useful
comments including the Fourier power spectrum (shown in Fig.~5).


\begin{thebibliography}{50}
\expandafter\ifx\csname natexlab\endcsname\relax\def\natexlab#1{#1}\fi

\bibitem[{{Athanassoula}(1984)}]{Athanassoula84}
{Athanassoula}, E. 1984, \physrep, 114, 321

\bibitem[{{Bertin}(1983)}]{Bertin1983}
{Bertin}, G. 1983, in IAU Symposium, Vol. 100, Internal Kinematics and Dynamics
  of Galaxies, ed. E.~{Athanassoula}, 119

\bibitem[{{Bertin}(1993)}]{Bertin1993}
{Bertin}, G. 1993, \pasp, 105, 640

\bibitem[{{Bertin} \& {Lin}(1996)}]{Bertin96}
{Bertin}, G., \& {Lin}, C.~C. 1996, {Spiral structure in galaxies a density
  wave theory} (Cambridge, MA MIT Press, 1996 Physical description x, 271
  p.~ISBN0262023962), ISBN0262023962

\bibitem[{{Bertin} {et~al.}(1989{\natexlab{a}}){Bertin}, {Lin}, {Lowe}, \&
  {Thurstans}}]{Bertin1989a}
{Bertin}, G., {Lin}, C.~C., {Lowe}, S.~A., \& {Thurstans}, R.~P.
  1989{\natexlab{a}}, \apj, 338, 78

\bibitem[{{Bertin} {et~al.}(1989{\natexlab{b}}){Bertin}, {Lin}, {Lowe}, \&
  {Thurstans}}]{Bertin1989b}
---. 1989{\natexlab{b}}, \apj, 338, 104

\bibitem[{{Binney} \& {Tremaine}(1987)}]{BT1987}
{Binney}, J., \& {Tremaine}, S. 1987, {Galactic dynamics} (Princeton, NJ,
  Princeton University Press, 1987, 747 p.)

\bibitem[{{Carlberg} \& {Sellwood}(1985)}]{Carlberg85}
{Carlberg}, R.~G., \& {Sellwood}, J.~A. 1985, \apj, 292, 79

\bibitem[{{D'Onghia} {et~al.}(2013){D'Onghia}, {Vogelsberger}, \&
  {Hernquist}}]{DOnghiaetal2013}
{D'Onghia}, E., {Vogelsberger}, M., \& {Hernquist}, L. 2013, \apj, 766, 34

\bibitem[{{Elmegreen} \& {Elmegreen}(1983)}]{EE1983}
{Elmegreen}, B.~G., \& {Elmegreen}, D.~M. 1983, \apj, 267, 31

\bibitem[{{Elmegreen} {et~al.}(1992){Elmegreen}, {Elmegreen}, \&
  {Montenegro}}]{Elmegreen1992}
{Elmegreen}, B.~G., {Elmegreen}, D.~M., \& {Montenegro}, L. 1992, \apjs, 79, 37

\bibitem[{{Elmegreen} {et~al.}(1989){Elmegreen}, {Seiden}, \&
  {Elmegreen}}]{Elmegreen1989}
{Elmegreen}, B.~G., {Seiden}, P.~E., \& {Elmegreen}, D.~M. 1989, \apj, 343, 602

\bibitem[{{Elmegreen} \& {Thomasson}(1993)}]{Elmegreen1993}
{Elmegreen}, B.~G., \& {Thomasson}, M. 1993, \aap, 272, 37

\bibitem[{{Goldreich} \& {Lynden-Bell}(1965)}]{GoldreichLyndenBell65}
{Goldreich}, P., \& {Lynden-Bell}, D. 1965, \mnras, 130, 125

\bibitem[{{Grosbol} \& {Patsis}(1998)}]{GrosbolPatsis1998}
{Grosbol}, P.~J., \& {Patsis}, P.~A. 1998, \aap, 336, 840

\bibitem[{{Huber} \& {Pfenniger}(2002)}]{Huber02}
{Huber}, D., \& {Pfenniger}, D. 2002, \aap, 386, 359

\bibitem[{{Jenkins} \& {Binney}(1990)}]{JenkinsBinney1990}
{Jenkins}, A., \& {Binney}, J. 1990, \mnras, 245, 305

\bibitem[{{Kuijken} \& {Dubinski}(1995)}]{KD1995}
{Kuijken}, K., \& {Dubinski}, J. 1995, \mnras, 277, 1341

\bibitem[{{Kuijken} \& {Gilmore}(1991)}]{KuijkenGilmore1991}
{Kuijken}, K., \& {Gilmore}, G. 1991, \apjl, 367, L9

\bibitem[{{Lewis} \& {Freeman}(1989)}]{LewisFreeman1989}
{Lewis}, J.~R., \& {Freeman}, K.~C. 1989, \aj, 97, 139

\bibitem[{{Lin}(1970)}]{Lin70}
{Lin}, C.~C. 1970, in IAU Symposium, Vol.~38, The Spiral Structure of our
  Galaxy, ed. W.~{Becker} \& G.~I. {Kontopoulos}, 377

\bibitem[{{Lin} \& {Bertin}(1985)}]{LinBertin1985}
{Lin}, C.~C., \& {Bertin}, G. 1985, in IAU Symposium, Vol. 106, The Milky Way
  Galaxy, ed. H.~{van Woerden}, R.~J. {Allen}, \& W.~B. {Burton}, 513--530

\bibitem[{{Lin} \& {Shu}(1964)}]{LinShu1964}
{Lin}, C.~C., \& {Shu}, F.~H. 1964, \apj, 140, 646

\bibitem[{{Lindblad}(1962)}]{Lindblad1962}
{Lindblad}, B. 1962, in IAU Symposium, Vol.~15, Problems of Extra-Galactic
  Research, ed. G.~C. {McVittie}, 146

\bibitem[{{Lindblad}(1963)}]{Lindblad1963}
{Lindblad}, B. 1963, Stockholms Observatoriums Annaler, 22

\bibitem[{{Lowe} {et~al.}(1994){Lowe}, {Roberts}, {Yang}, {Bertin}, \&
  {Lin}}]{Lowe1994}
{Lowe}, S.~A., {Roberts}, W.~W., {Yang}, J., {Bertin}, G., \& {Lin}, C.~C.
  1994, \apj, 427, 184

\bibitem[{{Lynden-Bell} \& {Kalnajs}(1972)}]{Lynden-BellKalnajs1972}
{Lynden-Bell}, D., \& {Kalnajs}, A.~J. 1972, \mnras, 157, 1

\bibitem[{{Mark}(1976{\natexlab{a}})}]{Mark1976b}
{Mark}, J.~W.-K. 1976{\natexlab{a}}, \apj, 203, 81

\bibitem[{{Mark}(1976{\natexlab{b}})}]{Mark1976a}
{Mark}, J.~W.~K. 1976{\natexlab{b}}, \apj, 205, 363

\bibitem[{{Mark}(1976{\natexlab{c}})}]{Mark1976c}
{Mark}, J.~W.-K. 1976{\natexlab{c}}, \apj, 206, 418

\bibitem[{{Mark}(1977)}]{Mark1977}
---. 1977, \apj, 212, 645

\bibitem[{{Masset} \& {Tagger}(1997)}]{Masset1997}
{Masset}, F., \& {Tagger}, M. 1997, \aap, 322, 442

\bibitem[{{McMillan} \& {Dehnen}(2007)}]{McMillan2007}
{McMillan}, P.~J., \& {Dehnen}, W. 2007, \mnras, 378, 541

\bibitem[{{Puerari} {et~al.}(2000){Puerari}, {Block}, {Elmegreen}, {Frogel}, \&
  {Eskridge}}]{Puerari2000}
{Puerari}, I., {Block}, D.~L., {Elmegreen}, B.~G., {Frogel}, J.~A., \&
  {Eskridge}, P.~B. 2000, \aap, 359, 932

\bibitem[{{Ro{\v s}kar} {et~al.}(2012){Ro{\v s}kar}, {Debattista}, {Quinn}, \&
  {Wadsley}}]{Roskeretal2012}
{Ro{\v s}kar}, R., {Debattista}, V.~P., {Quinn}, T.~R., \& {Wadsley}, J. 2012,
  \mnras, 426, 2089

\bibitem[{{Saha} {et~al.}(2012){Saha}, {Martinez-Valpuesta}, \&
  {Gerhard}}]{sahaetal2012}
{Saha}, K., {Martinez-Valpuesta}, I., \& {Gerhard}, O. 2012, \mnras, 421, 333

\bibitem[{{Saha} {et~al.}(2010){Saha}, {Tseng}, \& {Taam}}]{Sahaetal2010}
{Saha}, K., {Tseng}, Y., \& {Taam}, R.~E. 2010, \apj, 721, 1878

\bibitem[{{Salo} \& {Laurikainen}(2000)}]{Salo00}
{Salo}, H., \& {Laurikainen}, E. 2000, \mnras, 319, 393

\bibitem[{{Salo} {et~al.}(2010){Salo}, {Laurikainen}, {Buta}, \&
  {Knapen}}]{Salo10}
{Salo}, H., {Laurikainen}, E., {Buta}, R., \& {Knapen}, J.~H. 2010, \apjl, 715,
  L56

\bibitem[{{Sellwood}(1985)}]{Sellwood1985}
{Sellwood}, J.~A. 1985, \mnras, 217, 127

\bibitem[{{Sellwood}(2011)}]{Sellwood2011}
---. 2011, \mnras, 410, 1637

\bibitem[{{Sellwood}(2012)}]{Sellwood2012}
---. 2012, \apj, 751, 44

\bibitem[{{Sellwood} \& {Binney}(2002)}]{SellwoodBinney2002}
{Sellwood}, J.~A., \& {Binney}, J.~J. 2002, \mnras, 336, 785

\bibitem[{{Sellwood} \& {Carlberg}(2014)}]{Sellwood2014}
{Sellwood}, J.~A., \& {Carlberg}, R.~G. 2014, \apj, 785, 137

\bibitem[{{Springel} {et~al.}(2001){Springel}, {Yoshida}, \&
  {White}}]{Springeletal2001}
{Springel}, V., {Yoshida}, N., \& {White}, S.~D.~M. 2001, \na, 6, 79

\bibitem[{{Sygnet} {et~al.}(1988){Sygnet}, {Tagger}, {Athanassoula}, \&
  {Pellat}}]{Sygnet1988}
{Sygnet}, J.~F., {Tagger}, M., {Athanassoula}, E., \& {Pellat}, R. 1988,
  \mnras, 232, 733

\bibitem[{{Toomre}(1969)}]{Toomre1969}
{Toomre}, A. 1969, \apj, 158, 899

\bibitem[{{Toomre}(1981)}]{Toomre1981}
{Toomre}, A. 1981, in Structure and Evolution of Normal Galaxies, ed.
  {S.~M.~Fall \& D.~Lynden-Bell}, 111--136

\bibitem[{{Toomre} \& {Kalnajs}(1991)}]{ToomreKalnajs91}
{Toomre}, A., \& {Kalnajs}, A.~J. 1991, in Dynamics of Disc Galaxies, ed.
  B.~{Sundelius}, 341

\bibitem[{{Toomre} \& {Toomre}(1972)}]{ToomreToomre72}
{Toomre}, A., \& {Toomre}, J. 1972, \apj, 178, 623

\end{thebibliography}
\end{document}